# Directive Emission Obtained by Coordinate Transformation

Jingjing Zhang [1], Yu Luo [1], Hongsheng Chen [1,2*], Lixin Ran [1], Bae-Ian Wu [2], and Jin Au Kong [1,2]

[1] *The Electromagnetics Academy at Zhejiang University, Zhejiang University, Hangzhou 310058, P. R. China*

[2] *Research Laboratory of Electronics, Massachusetts Institute of Technology, Cambridge, Massachusetts 02139*

We use coordinate transformation theory to realize substrates that can modify the emission of an embedded source. Simulation results show that with proper transformation functions the energy radiated by a source embedded in these space variant media will be concentrated in a narrow beam. The thickness of the slab achieved with our transformations will no longer be restricted by the evanescent modes and the source can be placed at any position along the boundary of the substrate without affecting the radiation pattern. We also discuss the case where reduced parameters are used, which still performs well and is physically realizable.

[*] Author to whom correspondence should be addressed; electronic mail: hansomchen@zju.edu.cn

Based on the coordinate transformation theory, J. B. Pendry introduced the idea of cloak which could conceal a volume to detectors by wrapping it with an appropriately designed coating [1]. Following this approach, a microwave invisibility cloak with simplified constitutive parameters was proposed and experimentally realized using metamaterial [2]. Much more attentions are given afterwards [3-11] and some other novel devices, such as the EM concentrators [12, 13], rotators [14], and hyperlens [15, 16] were also investigated by similar methods. These rapid progresses make the coordinate transformation approach a hot topic in electromagnetics and imply very important future applications [17, 18]. In this paper, we propose the method of utilizing coordinate transformation to obtain directive emission. Simulation results show the energy radiated by a source embedded in these space variant planar media will be concentrated in a narrow beam. In contrast to some traditional metamaterial substrate antenna [19, 20], thickness of this planar antenna achieved with transformation media will no longer be restricted by the evanescent modes, and the feeding source can be placed at any position along the boundary of the substrate without affecting the radiation pattern. Simplified parameters have also been proposed, which indicate the possibility of physical realization.

We firstly discuss two dimensional cases. Consider the transformation between the initial cylindrical coordinate ($\rho', \varphi', z'$) and the transformed Cartesian coordinate ($x, y, z$)

$$x = \frac{d}{R}\rho', \quad y = l\sin\varphi', \quad z = z'. \tag{1}$$

where $R$ is the radius of a cylinder in the initial coordinate. $d$ and $2l$ represent the thickness and width of the slab in the transformed coordinate respectively, as shown in Fig. 1

If the initial virtual domain is free space characterized by constitutive parameters $\varepsilon_0$ and $\mu_0$, the permittivity and permeability tensors of the transformation media can be expressed as [4, 5]

$$\bar{\bar{\varepsilon}} = \varepsilon_0 \left( \frac{x}{\sqrt{l^2-y^2}} \hat{x}\hat{x} + \frac{\sqrt{l^2-y^2}}{x} \hat{y}\hat{y} + \left(\frac{R}{d}\right)^2 \frac{x}{\sqrt{l^2-y^2}} \hat{z}\hat{z} \right)$$
$$\bar{\bar{\mu}} = \mu_0 \left( \frac{x}{\sqrt{l^2-y^2}} \hat{x}\hat{x} + \frac{\sqrt{l^2-y^2}}{x} \hat{y}\hat{y} + \left(\frac{R}{d}\right)^2 \frac{x}{\sqrt{l^2-y^2}} \hat{z}\hat{z} \right)$$

(2)

Note that the transformation described by equations (1) is conformal. Under this conformal mapping, a cylindrical region ($0 \leq \rho' \leq R$) has been transformed to a slab ($0 \leq x \leq d$, $-l \leq y \leq l$), as depicted in Fig. 1. We can see that the origin ($\rho'=0$) is mapped to the line $x=0$, and the blue and red curves indicate the corresponding mapping in the original and transformed spaces. It should be noted that the curve $\rho'=R$ has been transformed to the line $x=d$. We know that a line source in free space will naturally radiate cylindrical wave, which means the wave vector $k$ is always along radial direction. Due to the conformality of this transformation, the EM waves radiated from a line source located at any position along the left boundary ($x=0$) will always propagate along x direction (perpendicular to the boundary marked in red). What we need to point out is that although this transformation will cause the EM fields mismatched at the right boundary ($x=d$), the wave transmitted from the slab is still a plane wave because the wave vector $k$ is always perpendicular to the right boundary (the diffraction can be ignored if the slab is sufficiently large $l \gg \lambda$). The mismatch will only reduce the enhancement at the emitting direction, but will not disturb the directivity.

The performance of the slab antenna is simulated and validated by the finite element method. The simulation quantities are normalized to unity and all domain boundaries are assumed as perfectly matched layers in order to prevent reflections. In all the following simulations, the width $l$ of the slab is 1 $m$ and we use a TE polarized line source to excite the wave at 2GHz. Fig. 2 shows the $E_z$ distribution when the source is embedded at the left boundary of the slab. In Fig. 2(a) the source is located at the center of the left boundary of the slab which is 0.1 $m$ thick. As the thickness of the slab is reduced to 0.05 $m$, which is only 1/3 of the wavelength, the electric field distribution remains similar as shown in Fig. 2 (b). Fig. 2 (c) is the case when source is fed 0.3 $m$ from the center along the left boundary. The simulation results show that reducing the thickness of the slab and changing the location of the source along the boundary will both have little effect on directivity, but the magnitude of the emitted EM fields is mitigated when the source is not located at the center. This is because the deviation of the source to the center will cause more mismatch of the field at the boundary, which will result in more reflections. The corresponding far field power density in terms of angle $\varphi$ of the three cases is plotted in Fig. 3 (a). Compared with the traditional metamaterial antenna of which the thickness should be sufficiently large so that the evanescent modes can be attenuated in the slab and negligible on the interface [19, 20, 21], the overlapping of the black solid line and red dashed line confirms this antenna does not have this thickness restriction since the wave vector $k$ is naturally perpendicular to the interface as discussed above. The blue dash dot line in Fig.3 (a) indicates the directivity of the antenna is also insensitive to the position of the source, and a source fed asymmetrically along the left boundary will only cause a suppression of the enhancement. The far field radiation pattern of the source embedded at the center of the left boundary when using 0.05 $m$ thick slab is shown in Fig .3(b). The half-power beamwidth is about 4.9° from the simulation.

Notice that in Equations (2), the three components of $\varepsilon$ are all spatially varied

(so as $\mu$). And the y components of $\varepsilon$ and $\mu$ are infinite at the origin while the other four parameters are infinite at the edge of $y = l$. This increases the difficulty in fabrication and hence we need to figure out ways to simplify the parameters in order to make it possible to realize.

When the line source is TE polarized along z direction, only $\varepsilon_z$, $\mu_x$ and $\mu_y$ in Equations (2) enter into Maxwell's equations. Moreover, the dispersion properties and wave trajectory in the slab remain the same, as long as the value $\varepsilon_z \mu_x$ and $\varepsilon_z \mu_y$ are kept constant. This gives the ability to choose one of the three constitutive parameters arbitrarily to achieve some favorable condition [2, 3, 6, 9]. One choice is to use the following reduced parameters

$$\varepsilon_z = \varepsilon_0 \left(\frac{R}{d}\right)^2, \quad \mu_x = \mu_0 \frac{x^2}{l^2 - y^2}, \quad \mu_y = \mu_0. \tag{3}$$

From the simulation result we find that the slab antenna with reduced parameters still has a good performance even if we use a very thin slab (0.05 *m* in the simulation), which is shown in Fig. 4. Fig. 5 (a) displays the normalized power density in the far field region of the two reduced cases. Fig. 5 (b) is the far field radiation pattern of a line source embedded at the center of left boundary when using a thin slab (0.05 *m*) with reduced parameters, indicating a half-power beamwidth of around 5.6°. This simulation results demonstrate that the medium with reduced parameters is a good choice for physical realization and provides a simpler way to an experimental demonstration.

It should be noted that this idea can also be extended to three-dimensional case to obtain high directive antenna in both **E** plane and **H** plane. Such mapping can be obtained by transforming a spherical wave to a plane wave. Consider the following transformation between the initial spherical coordinate ($r', \theta', \varphi'$) and the transformed

cylindrical coordinate ($\rho, \varphi, z$):

$$\rho = l\sin\theta', \quad \varphi = \varphi', \quad z = \frac{d}{R}r' \tag{4}$$

Under this conformal mapping, a sphere with radius $R$ has been transformed to a cylinder slab with radius $l$ and thickness $d$ ($0 \leq z \leq d, \rho \leq l$). The permittivity and permeability tensors of the transformation medium are given by

$$\bar{\bar{\varepsilon}} = \varepsilon_0 \left( \frac{\sqrt{l^2 - \rho^2}}{z} \hat{\rho}\hat{\rho} + \left(\frac{R}{d}\right)^2 \frac{z}{\sqrt{l^2 - \rho^2}} \hat{\varphi}\hat{\varphi} + \frac{z}{\sqrt{l^2 - \rho^2}} \hat{z}\hat{z} \right)$$

$$\bar{\bar{\mu}} = \mu_0 \left( \frac{\sqrt{l^2 - \rho^2}}{z} \hat{\rho}\hat{\rho} + \left(\frac{R}{d}\right)^2 \frac{z}{\sqrt{l^2 - \rho^2}} \hat{\varphi}\hat{\varphi} + \frac{z}{\sqrt{l^2 - \rho^2}} \hat{z}\hat{z} \right) \tag{5}$$

In this condition, the EM waves radiated from a dipole located close to the boundary $z = 0$ of the slab will be concentrated in a narrow cone in the z direction. Since the phenomenon is very similar to the 2D case, we do not elaborate 3D case here.

In conclusion, a new approach to realize high directive antenna using coordinate transformation is proposed. Both two dimensional and three dimensional cases are discussed, and the reduced parameters under two-dimensional condition are also given, which makes the realization of this antenna a practical possibility. The high directive radiation pattern of the antenna designed in this way is insensitive to the positions of the source. Such functionality of the directive antenna is successfully demonstrated by numerical simulations, which also provides a good application of the transformation method.

This work is sponsored by the Chinese National Science Foundation under Grant Nos. 60531020 and 60671003, the NCET-07-0750, the China Postdoctoral Science Foundation under Grant No. 20060390331, the ONR under Contract No. N00014-01-1-0713, and the Department of the Air Force under Air Force Contract No.F19628-00-C-0002.

FIG. 1 (Color online) Spatial coordinate transformation for the design of the high directive slab antenna. (a) Original space. (b) Transformed space.

FIG. 2 (Color online) $E_z$ distribution when the antenna is fed by $z$-polarized TE wave. (a) The thickness of the slab is 0.1 *m* and the source is embedded at the center of the left boundary of the slab. (b) The thickness of the slab is 0.05 *m* and the source is embedded at the center of the left boundary of the slab. (c) The thickness of the slab is 0.1 m and the source is embedded at the point 0.3 *m* from the center of the left boundary of the slab.

FIG. 3 (Color online) (a) Normalized power density in the far field region of the three cases. The black solid line, red dashed line and blue dashed dot line correspond to the embedded source fed at the center of the left boundary of 0.1 *m* thick slab, 0.05 *m* thick slab, 0.3 *m* to the center of the left boundary of 0.1 *m* thick slab respectively (b) Far field pattern for the case when the source is embedded at the center of the left boundary of 0.05 *m* thick slab.

FIG. 4 (Color online) $E_z$ distribution when the antenna with reduced parameters is fed by $z$-polarized TE wave. (a) The thickness of the slab is 0.1 *m* and the source is embedded at the center of the left boundary of the slab. (b) The thickness of the slab is 0.05 *m* and the source is embedded at the center of the left boundary of the slab.

FIG. 5 (Color online) Normalized power density in the far field region of the two reduced cases. The black solid line and red dashed line correspond to the embedded source fed at the center of the left boundary of 0.1 *m* thick slab, 0.05 *m* thick slab

respectively. (b) Far field pattern for the case where the source is embedded at the center of the left boundary of 0.05 *m* thick slab.

FIG. 1

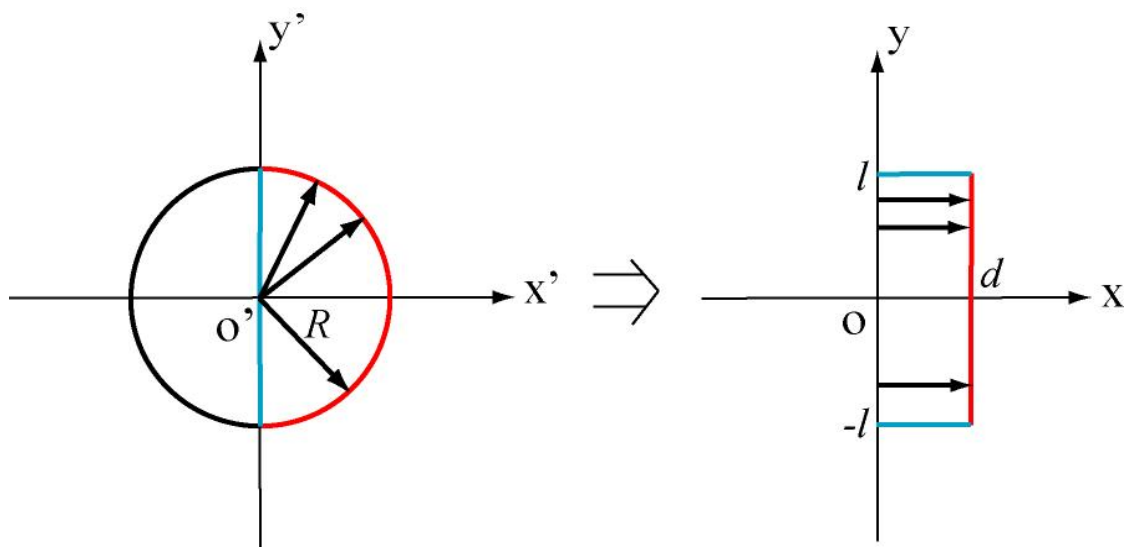

FIG. 2

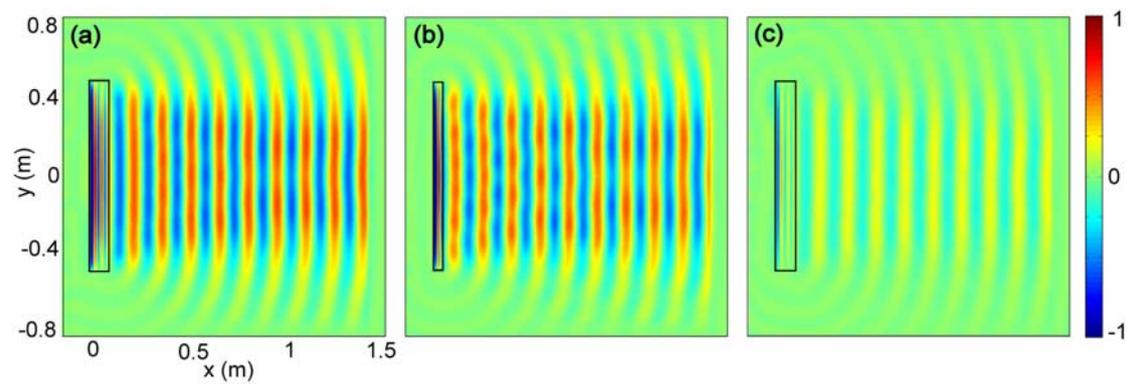

FIG. 3

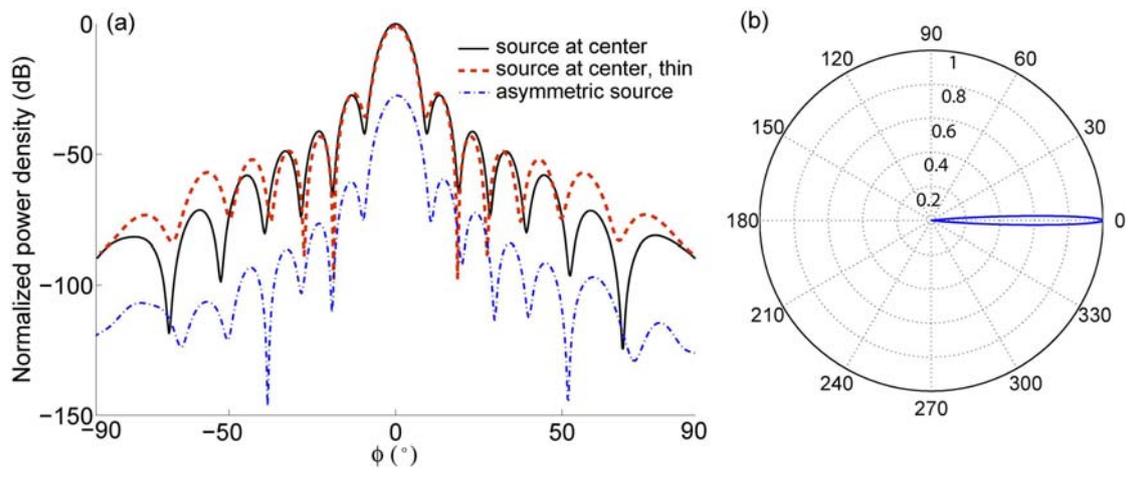

FIG. 4

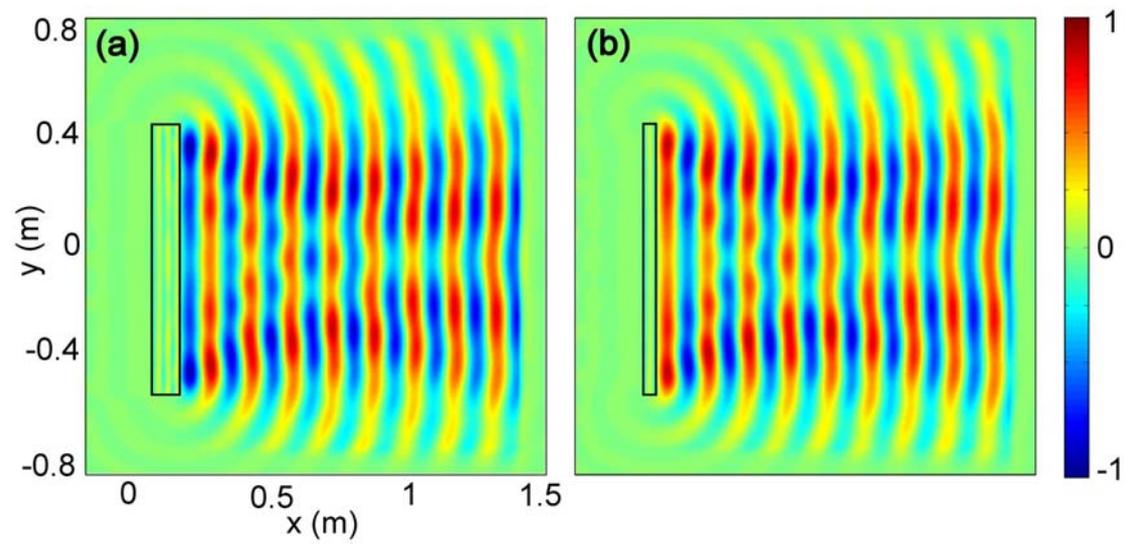

FIG. 5

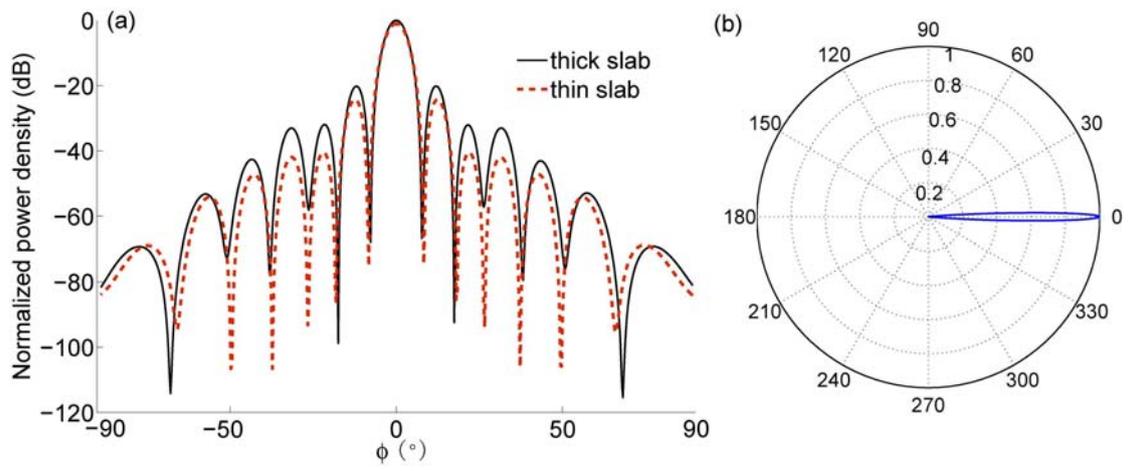